\documentclass[12pt]{article}
\usepackage{graphics,color}
\begin{document}
\thispagestyle{empty}
\noindent\
\\
\\
\\
\begin{center}
\large \bf Composite Weak Bosons and Dark Matter
\end{center}
\hfill
 \vspace*{1cm}
\noindent
\begin{center}
{\bf Harald Fritzsch}\\
Department f\"ur Physik\\ 
Ludwig-Maximilians-Universit\"at\\
M\"unchen, Germany \\
\end{center}

\begin{abstract}

The three weak bosons are bound states of two fermions and their antiparticles. There exist also two bound states of three fermions. One of them is neutral and stable. This particle provides the dark matter in our universe

\end{abstract}

\newpage

In the Standard Theory of particle physics the massive weak bosons are pointlike and obtain their masses by the 
spontaneous breaking of the gauge symmetry. I do not like this kind of mass generation. The concept of mass is a fundamental concept and there should be a dynamical mechanism to generate all masses in physics, the proton mass, the masses of the leptons and quarks and the masses of the weak bosons. \\

The masses of the bound states in confining gauge theories are generated by a mechanism, which is due to the confinement property  of the non-Abelean gauge bosons. For example, in the theory of Quantum Chromodynamics the masses of the hadrons can be calculated. They are given  by the field energy of the confined gluons and quarks and are proportional to the QCD mass scale $\Lambda_c$, which has  to be measurend in the experiments: $\Lambda_c = 332 \pm 17 $  MeV.\\

In the Standard Theory the physics of the six leptons and the six quarks depends on 20 free prameteres: twelve masses of the leptons and quarks and eight parameters of the flavor mixing - six mixing angles and two phase parameters, which describe the CP violation.\\

These twenty parameters might indicate, that the leptons, quarks and weak bosons are not pointlike, but composite systems. Especially we shall assume, that inside the leptons and quarks are two pointlike particles, a fermion and  a scalar. There are two fermions and four scalars (see also ref. (1)). They are called "haplons" ("haplos" means "simple" in the Greek language).\\

The haplons are confined by massless gauge bosons, which are described by a gauge theory similar to Quantum Chromodynamics.  We call these gauge bosons "hypergluons" and the associated quantum numbers "hypercolors". The new gauge theory is called "Quantum Haplodynamics" (QHD). I shall not discuss the internal structure of the leptons and quarks in more detail, but come now to the weak bosons. \\

We assume that the masses of the weak bosons are also generated dynamically. This is possible, if the weak bosons are not elementary gauge bosons, but bound states of haplons, analogous to the $\rho$-mesons in QCD.\\

The weak bosons consist of two haplons, a fermion and its antiparticle (see also ref.(2,3,4,5,6)). The $QHD$ mass scale is described by a mass parameter $\Lambda_h$, which determines in particular the size of the weak bosons.\\ 

The haplons are massless and interact with each other through the exchange of massless hypergluons. The number of hypergluons depends on the gauge group, which is unknown. We assume, that it is the same as the gauge group of $QCD$: $SU(3)$. Thus the binding of the haplons is due to the exchange of eight massless hypergluons. \\ 

Two types of fermions are needed as constituents of the weak bosons, denoted by $\alpha$ and $\beta$. The electric charges of these two haplons are (+2/3) and (-1/3). Before the mixing of the neutral weak boson and the photon the three weak bosons have the following internal structure:

\begin{eqnarray}
W^+ = (\overline{\beta} \alpha),~~ W^- = (\overline{\alpha} \beta),~~W^3 =\frac{1}{\sqrt{2}} \left( \overline{\alpha} \alpha - \overline{\beta} \beta \right).
\end{eqnarray}

The $QHD$ mass scale can be estimated, using the observed value of the decay constant of the weak boson. This constant is analogous to the decay constant of the $\rho$-meson, which is directly related to the $QCD$ mass scale. The decay constant of the weak boson is given by the mass of the weak boson and the observed value of the weak mixing angle:

\begin{eqnarray}
F_{\rm W}=\sin\theta_{\rm W}\cdot \frac{M_{\rm W}}{e}.
\end{eqnarray}

We find $F_{\rm W} \simeq 0.124 ~{\rm TeV}$. Thus we expect, that the $QHD$ mass scale $\Lambda_h$ is  in the range between 0.2 TeV and 0.6 TeV. The ratio of the $QHD$ mass scale and of the $QCD$ mass scale is about three 
orders of magnitude.\\

There should exist also fermions, which consist of three haplons. The ground states are two particles: 

\begin{eqnarray}
(\alpha\alpha\beta)=Y^+ , (\alpha\beta\beta)=Y^0.
\end{eqnarray}

If the electromagnetic interaction is neglected, these two fermions would have the same mass. After introduction of 
electromagnetism this degeneracy is lifted. The electromagnetic field around the charged Y-particle contributes to the mass. We have in particular:  

\begin{eqnarray}
m(Y^+) =m(Y^0) + \delta m.
\end{eqnarray}

Let us assume, that the mass of the $Y^0$ - particle is 300 GeV. In this case the electromagnetic massshift $\delta m$ is about 200 MeV:\\

\begin{eqnarray}
m(Y^+)  \simeq 300.2~{\rm GeV}.
\end{eqnarray}

The charged Y - particle decays into the neutral one via the weak interaction, for example $Y^+ \to Y^0 + \pi^+$.\\

The lifetime of the charged Y - particle is about $10^{-9}$ seconds. The neutral Y-particle consists of three haplons. It is stable, since the haplon number is conserved.\\ 

In proton-proton or antiproton-proton collisions the Y-particles must be produced in pairs, ( e.g.  $p + p \to Y^+ \overline{ Y^+}$ + ...). We consider now the production of a  $Y^{+}$ - particle and its antiparticle. This particle decays into a $ Y^0$ - particle by emitting a virtual positively charged weak boson:

\begin{eqnarray}
Y^{+} \to Y^0 +``W^+" 
\end{eqnarray}

For example, the charged Y-particle can decay into the neutral one by emitting a charged pion. The $Y^+$ - particle decays inside the detector of the LHC. The track of this particle can probably be observed. The length of the track might be 
0,5 m.\\ 

The $Y^0$ - particle or its antiparticle cannot be directly observed, but indirectly due to the missing energy. In such an event one could also observe the decay products of the two virtual weak bosons, for example two pions. \\

During the Big Bang the Y-fermions and their antiparticles would be produced, as well as quarks and antiquarks or leptons and their antiparticles. Since there is a small asymmetry between the matter and the antimatter in the universe, due to the violation of the CP-symmetry, there are more quarks than antiquarks. The antiquarks disappear - they annihilate with the quarks and produce photons. The remaining quarks are the constituents of the matter in our universe.\\

Due to CP-violation there are also more $Y^0$-particles than $\overline{Y^0}$ - particles, or more $\overline{Y^0}$ - particles than  $Y^0$-particles. Those would annihilate with the $Y^0$ - particles or $\overline{Y^0}$ - particles, and finally the univserse would contain, besides protons and neutrons, a gas of stable $Y^0$ - particles or  $\overline{Y^0}$ - particles, providing the dark matter in our universe.\\

It is not possible to calculate, how much dark matter exists - this would imply that one can calculate the density of the $Y^0$ - particles or $\overline{Y^0}$ - particles, which is not possible at this time.\\

The average density of the dark matter in our galaxy has been measured to about $0.39 ~{\rm GeV}/{\rm cm}^3$.  If the mass of the neutral Y-particle is 300 GeV, there would be about 1 300 $Y^0$-particles or $\overline{Y^0}$ - particles in one cubic meter of space. Our galaxy is inside a big cloud of dark matter. In this cloud the neutral Y-particles move with an average speed of about 15 m/sec.\\

The sun moves around the center of our galaxy with a speed of about 200 km/s. The speed of the earth around the sun 
is about 31 km/s. The speed of the earth inside the dark matter cloud varies between 169 km/s and 231 km/s. Thus the Y-particles have  on earth a rather large speed.\\

A neutral Y-fermion can emit a virtual Z-boson, which interacts with an atomic nucleus. The cross section for the reaction of a neutral Y - fermion with a nucleus should be about $10^{-45}~{\rm cm}^2$. In a collision of a neutral Y-fermion with a nucleus one can observe in specific experiments the sudden change of the momentum of the nucleus. Thus the neutral Y-fermions can be observed indirectly, e.g. in the experiments in the Gran Sasso Laboratory, but the relevant cross section might be too small.\\ 

With the Large Hadron Collider the Y - particles are produced by the direct interaction of a quark and an antiquark. They cannot be produced by the interaction of two gluons. Thus the production cross section would be larger, if the Large Hadron Collider would be a proton-antiproton collider. But even with the present Large Hadron Collider I expect that the charged Y-fermions can be observed soon. \\

\end{document}